\documentstyle[prd,aps,floats]{revtex}

\def\be{\begin{equation}}
\def\ee{\end{equation}}
\def\bea{\begin{eqnarray}}
\def\eea{\end{eqnarray}}

\begin{document}
\preprint{PU-ICG/02-11, gr-qc/0206085} \draft

%
%
\input epsf
\renewcommand{\topfraction}{0.99}
\renewcommand{\bottomfraction}{0.99}
\twocolumn[\hsize\textwidth\columnwidth\hsize\csname
@twocolumnfalse\endcsname

\title{Cosmology with positive and negative exponential
  potentials}
\author{Imogen P.~C.~Heard and David Wands}
\address{Institute of Cosmology and Gravitation, University of
Portsmouth, Portsmouth PO1 2EG,~~~U.~K.}
\date{\today}
\maketitle
\begin{abstract}
  We present a phase-plane analysis of cosmologies containing a scalar
  field $\phi$ with an exponential potential $V \propto \exp(-\lambda
  \kappa \phi)$ where $\kappa^2 = 8\pi G$ and $V$ may be positive or
  negative. We show that power-law kinetic-potential scaling solutions
  only exist for sufficiently flat ($\lambda^2<6$) positive potentials
  or steep ($\lambda^2>6$) negative potentials. The latter correspond
  to a class of ever-expanding cosmologies with negative potential.
  However we show that these expanding solutions with a negative
  potential are to unstable in the presence of ordinary matter,
  spatial curvature or anisotropic shear, and generic solutions always
  recollapse to a singularity. Power-law kinetic-potential scaling
  solutions are the late-time attractor in a collapsing universe for
  steep negative potentials (the ekpyrotic scenario) and stable
  against matter, curvature or shear perturbations. Otherwise
  kinetic-dominated solutions are the attractor during collapse (the
  pre big bang scenario) and are only marginally stable with respect
  to anisotropic shear.
\end{abstract}

\pacs{PACS numbers: 98.80.Cq\\
Preprint PU-ICG/02-11, gr-qc/0206085}

\vskip2pc]

\section{Introduction}

Scalar fields are ubiquitous in modern cosmology whether part of the
matter sector, such as the Higgs field, or dilaton or moduli fields in
the gravitational sector. They extend the range of qualitative
behaviour found in homogeneous and isotropic
Friedmann-Robertson-Walker (FRW) models compared with conventional
perfect fluid models. Scalar fields with positive potential energy can
drive an accelerated expansion and thus provide possible models for
inflation in the early universe or dark energy at the present epoch.

In this paper we will consider models where the scalar field, $\phi$,
has a simple exponential potential
\begin{equation}
V = V_0 \exp (-\lambda\kappa\phi) \,,
\end{equation}
where $\kappa^2\equiv 8\pi G_N$ is the gravitational coupling and
$\lambda$ is a dimensionless constant characterising the slope of the
potential. 
Such potentials arise naturally in supergravity models~\cite{Townsend}.
The possible cosmological role of exponential potentials
has been investigated before as a means of driving a period of early
inflation~\cite{LM85,pplane,texas,assist}, ekpyrotic or pre big bang
collapse~\cite{ekpyrotic,pyro,FinBra,assistcon} or
quintessence~\cite{Wetterich,FJ,Amendola,HW}.  Although a period of
accelerated expansion requires $V>0$ we will also consider the case
where $V<0$.  Negative potentials have recently been considered both
in an expanding universe \cite{Linde} and in a collapsing phase
\cite{ekpyrotic,pyro,FinBra}.

The scale-invariant form makes the exponential potential particularly
simple to study analytically. There are well-known exact solutions
corresponding to power-law solutions for the cosmological scale factor
$a\propto t^p$ in a spatially flat FRW model~\cite{LM85}, but more
generally the coupled Einstein-Klein-Gordon equations for a single
field can be reduced to a one-dimensional system which makes it
particularly suited to a qualitative
analysis~\cite{pplane,texas,CLW98,vdH,vdH2,Amendola,HW}.
Here we adopt a system of dimensionless dynamical variables \cite{WE} which
have previously been used to study positive exponential potentials in
isotropic \cite{CLW98,vdH2} and anisotropic \cite{vdH} cosmologies, and
mostly recently the brane-world scenario~\cite{Dunsby}.

In Section II we review the known results for the one-dimensional
system with $V>0$ and extend the analysis to include negative
potentials. We then describe in section III the effect of introducing
an additional term in the Friedmann equation to describe the effect of
introducing a barotropic fluid, spatial curvature, or anisotropic
shear in a Bianchi type~I cosmology, thereby raising the
system to two-dimensional phase-plane.
We discuss the implications of our results for the stability of
particular solutions in section~IV, epsecially those solutions invoked
in the epyrotic and other pre big bang scenarios.

\section{One-dimensional phase-space}

We consider first a scalar field with an exponential potential energy
density evolving in a spatially-flat Friedmann--\-Robertson--\-Walker
(FRW) universe. The Klein-Gordon equation for the scalar field in an
FRW model is
\be
\ddot\phi + 3H\dot\phi + {dV\over d\phi} = 0 \ ,
\label{KG}
\ee
where the Hubble parameter, $H$, is determined by the Friedmann constraint
%
\be
\label{Friedmann}
H^2 = {\kappa^2\over3} \left( {1\over2}\dot\phi^2 + V \right) \ .
\ee

A homogeneous scalar field has energy density
$\rho_\phi=\dot\phi^2/2+V$ and pressure $P_\phi=\dot\phi^2/2-V$.
We introduce dimensionless variables~\cite{CLW98} \be
\label{defxy} x \equiv {\kappa\dot\phi \over \sqrt{6}\,H} \quad ;
\quad y \equiv {\kappa\sqrt{|V|} \over \sqrt{3}\,H} \ . \ee in
which the Friedmann constraint (\ref{Friedmann}) takes the simple
form \be \label{1Dcon} x^2 \pm y^2 = 1 \ . \ee
Throughout we
will use upper/lower signs to denote the two distinct cases of
$\pm V>0$. $x^2$ measures the contribution to the expansion
due to the field's kinetic energy density, while $\pm y^2$
represents the contribution of the potential energy. Their ratio
determines the equation of state
\begin{equation}
\frac{P_\phi}{\rho_\phi} = \frac{{1\over2}\dot\phi^2 - V}{{1\over2}\dot\phi^2 +
  V} = \frac{x^2 \mp y^2}{x^2 \pm y^2} \,.
\end{equation}
For a positive potential this which ranges between a stiff fluid with
$P_\phi=\rho_\phi$, when the kinetic energy dominates, and
$P_\phi=-\rho_\phi$, when the
potential dominates. For a negative potential we have an ultra-stiff
fluid with $P_\phi>\rho_\phi$.

We will restrict our discussion of the existence and stability of
critical points to the upper half plane $y\geq0$, i.e., expanding
cosmologies with $H>0$, but the trajectories are symmetrical under
time reversal, $H\to-H$, corresponding to reflection symmetry
$x\to x$ and $y\to-y$. Thus early time solutions in an expanding
universe are the same as to the late time limit in a collapsing
cosmology.
Without loss of generality we will consider only $\lambda\geq0$ as
the system is also symmetric under $\lambda\to-\lambda$ and
$x\to-x$.

The evolution equation (\ref{KG}) can then be written as an autonomous
system~\cite{WE,CLW98}: 
\bea
\label{eomx}
x' & = & -3x(1-x^2) \pm \lambda \sqrt{{3\over2}} y^2
 \ , \\
\label{eomy}
y' & = & xy \left( 3x - \lambda \sqrt{{3\over2}} \right) \ ,
\eea
where a prime denotes a derivative with respect to the logarithm of
the scale factor, $N\equiv\ln(a)$. The constraint (\ref{1Dcon}) then
reduces the system to a one-dimensional phase-space corresponding to
the unit circle for $V\geq0$ or hyperbola for $V\leq0$.

\subsection{Critical points}

Critical points correspond to fixed points $(x_i,y_i)$ where both
$x'=0$ and $y'=0$. These are self-similar solutions as dimensionless
quantities are invariant under time-translation $N\to N+\Delta N$. For
example we have $\dot{H}/H^2=$constant, where
\begin{equation}
\label{dotH1}
{\dot{H}\over H^2} = -3x^2 \,.
\end{equation}
Integrating this equation shows that all critical points, where
$x=x_i=$non-zero constant, correspond to a power-law solution for the
scale factor in terms of cosmic time:
\begin{equation}
\label{powerlaw}
a \propto |t|^p \,, \quad {\rm where}\ p = {1\over3x_i^2} \,.
\end{equation}

The system~(\ref{eomx}) and~(\ref{eomy}) has at most three fixed points:
\begin{itemize}
\item[$A_\pm$]
Two {\em kinetic-dominated} solutions exist for any form of the potential
with $y_{A}=0$ and $x_{A_+}=+1$ or $x_{A_-}=-1$. These are equivalent
to stiff-fluid dominated evolution with $a\propto t^{1/3}$
irrespective of the nature of the potential. This is the form of
cosmological evolution usually invoked approaching the singularity in
the Einstein conformal frame in the pre big bang scenario~\cite{pbb}.
\item[$B$]
A {\em potential-kinetic-scaling} solution exists for
$\pm(6-\lambda^2)>0$ with
\begin{equation}
x_B = {\lambda\over\sqrt{6}}  \,, \qquad y_B =
\sqrt{1-{\lambda^2\over6}}   \,.
\end{equation}
This solution only exists for sufficiently flat potentials
($\lambda^2<6$) for positive potentials, or steep potentials
($\lambda^2>6$) for negative potentials.

The power-law exponent, $p=2/\lambda^2$, depends on the slope of the
potential.  For positive potentials with $\lambda^2<2$ and $H>0$ this
solution corresponds to the well-known power-law inflation solutions
with an accelerating scale factor $\ddot{a}>0$~\cite{LM85}. For
negative exponential potential with $\lambda\gg1$ and $H<0$ this
yields the accelerated collapse with $p\ll1$ recently invoked in the
ekpyrotic scenario~\cite{ekpyrotic,pyro}.
\end{itemize}

\subsection{Stability}
\label{1Dstability}

\begin{itemize}
\item[$A_\pm$]
  Linear perturbations $x\to x+u$ about the points $x_{A_+}=+1$ and
  $x_{A_-}=-1$ have exponential behaviour $u\propto e^{mN}$ where
  $m_+=\sqrt{6}(\sqrt{6}-\lambda)$ and $m_-=\sqrt{6}(\sqrt{6}+\lambda)$
  respectively. Thus for positive $\lambda$, $x_{A_-}=-1$ is always
  unstable and $x_{A_+}=+1$ is stable for $\lambda^2>6$ but unstable
  for $\lambda^2<6$.
\item[$B$]
  Linear perturbations about the potential-kinetic scaling solution
  decay with eigenvalue $m=(\lambda^2-6)/2$ and this solution is thus
  always stable when this point exists for a positive potential, but
  unstable when this solution exists for a negative potential.
\end{itemize}

\subsection{Behaviour at infinity}

\begin{figure}[t]
\centering
\leavevmode\epsfysize=6cm \epsfbox{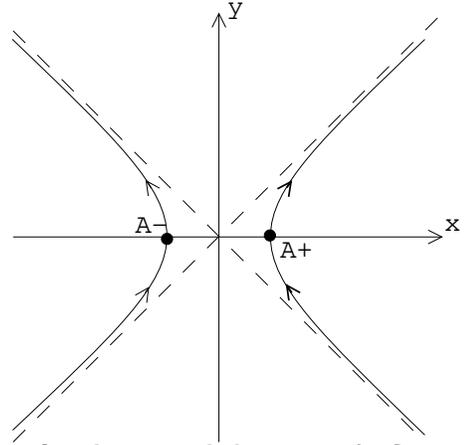}\\
\caption[hyperbflat]{\label{hyperbflat} One-dimensional phase-space
  for flat negative potentials, $\lambda^2<6$.}
\end{figure}

\begin{figure}[t]
\centering
\leavevmode\epsfysize=6cm \epsfbox{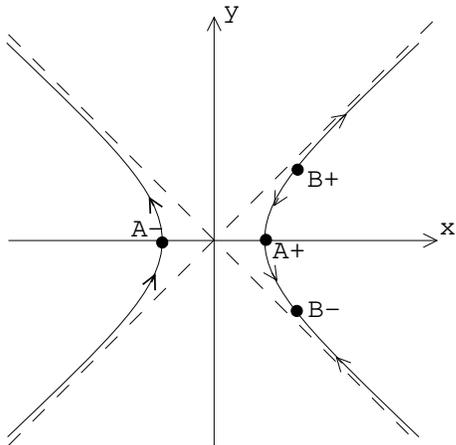}\\
\caption[hyperbsteep]{\label{hyperbsteep} One-dimensional phase-space
  for steep negative potentials, $\lambda^2>6$.}
\end{figure}

For positive potentials the one-dimensional system is compact, but for
$V\leq0$ the trajectories can go out to infinity with $x/y\to1$ and the
system reduces to
\be
x' \to 3x^3
\ee
so trajectories always approach infinity with $x^2= 1/6(N_*-N)$. 
Equation~(\ref{dotH1}) can then be integrated to obtain a solution for
$H$ and hence the scale factor in terms of proper time
\be
N_*-N \propto (t-t_*)^2 \,,
\ee
showing that the behaviour at infinity always corresponds to
recollapse, i.e., a maximum of the expansion at finite time $t_*$.
Thus expanding solutions in the upper half-plane are linked to
collapsing solutions in the lower half-plane at infinity. When $H$
changes sign, both $x$ and $y$ also change sign, so solutions in
Figures~\ref{hyperbflat} and~\ref{hyperbsteep} exiting
top-right are linked to solutions bottom-left and similarly top-left
to bottom-right.

\subsection{Qualitative evolution}

\paragraph{Flat positive potentials ($V>0$, $\lambda^2<6$)}. All three
critical points exist. Point $B$ is the stable late-time attractor and
points $A_-$ and $A_+$ are unstable repellors. Hence generic solutions
start in a kinetic-dominated regime and approach the kinetic-potential
scaling solution at late times.

\begin{figure}[t]
\centering
\leavevmode\epsfysize=6cm \epsfbox{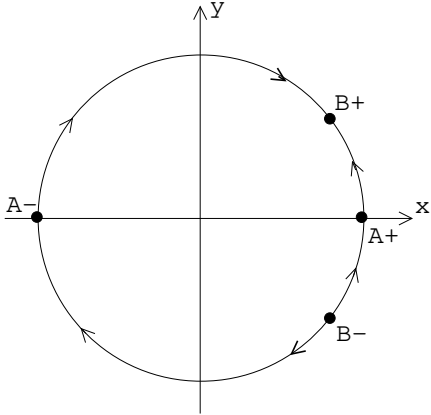}\\
\caption[circleflat]{\label{circleflat} One-dimensional phase-space
  for flat positive potentials, $\lambda^2<6$.
  Arrows indicate evolution in cosmic
  time, $t$. Note that in the lower half-plane, $H<0$, this has the opposite
  sense to $N\equiv\ln(a)$.}
\end{figure}

\paragraph{Steep positive potentials ($V>0$, $\lambda^2>6$)}. Only the two
kinetic-dominated fixed points $A_+$ and $A_-$ exist. For $\lambda>0$,
generic solutions start at $x_{A_-}=-1$ and approach $x_{A_+}=+1$ at
late times.

\begin{figure}[t]
\centering
\leavevmode\epsfysize=6cm \epsfbox{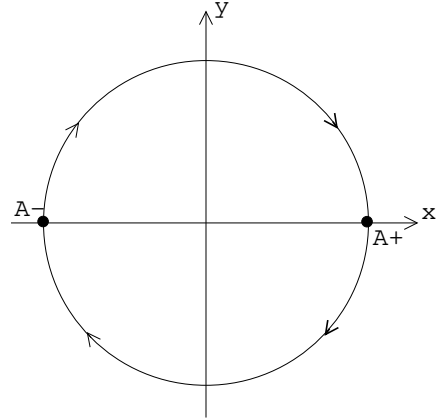}\\
\caption[circlesteep]{\label{circlesteep} One-dimensional phase-space
  for steep positive potentials, $\lambda^2>6$.}
\end{figure}

\paragraph{Flat negative potentials ($V<0$, $\lambda^2<6$)}. Only the two
kinetic-dominated fixed points $A_+$ and $A_-$ exist and both are
unstable repellors. Thus generic solutions begin in one of the kinetic
dominated solutions and go out to infinity, i.e., recollapse. Thus
these cosmological solutions have a finite lifetime, beginning in a
kinetic-dominated big bang (point $A_-$ or $A_+$), but recollapsing
after finite time and collapsing to a big crunch (at $A_+$ or $A_-$).

\paragraph{Steep negative potentials ($V<0$, $\lambda^2>6$)}. All three
critical points exist. For $\lambda>0$, point $A_+$ is the stable
late-time attractor. Thus left hyberbola describes solutions that
leave $x_{A_-}=-1$ and go to infinity. Generic solutions on the right
hyperbola start in the kinetic-potential scaling solution at point $B$
and then either become kinetic-dominated (approaching $A_{+}$) or go
to infinity (recollapse).

\begin{table*}[t]
\begin{center}
\begin{tabular}{||cc||ccc||}
Potential &  &  & Evolution &
 \\
\hline
\hline
Positive & flat & $A_-$ & $\rightarrow$ & $B_+$ \\
 & & big bang & expansion & future infinity \\
\hline
 & & $A_+$ & $\rightarrow$ & $B_+$ \\
 & & big bang & expansion & future infinity \\
\hline
 & & $B_-$ & $\rightarrow$ & $A_-$ \\
 & & past infinity & collapse & big crunch \\
\hline
 & & $B_-$ & $\rightarrow$ & $A_+$  \\
 & & past infinity & collapse & big crunch \\
\hline
 & steep & $A_-$ & $\rightarrow$ & $A_+$ \\
 & & big bang & expansion & future infinity \\
\hline
 & & $A_+$ & $\rightarrow$ & $A_-$ \\
 & & past infinity & collapse & big crunch \\
\hline
\hline
Negative & flat & $A_-$ & $\rightarrow$ & $A_+$ \\
 & & big bang & expansion and recollapse & big crunch \\
\hline
 & & $A_+$ & $\rightarrow$ & $A_-$ \\
 & & big bang & expansion and recollapse & big crunch \\
\hline
 & steep & $B_+$ & $\rightarrow$ & $A_+$ \\
 & & big bang & expansion & future infinity \\
\hline
 & & $B_+$ & $\rightarrow$ & $A_-$  \\
 & & big bang & expansion and recollapse & big crunch \\
\hline
 & & $A_-$ & $\rightarrow$ & $B_-$ \\
 & & big bang & expansion and recollapse & big crunch \\
\hline
 & & $A_+$ & $\rightarrow$ & $B_-$ \\
 & & past infinity & collapse & big crunch \\
\hline
\hline
\end{tabular}
\end{center}
\caption[tabl]{\label{tabl} The evolution of one-dimensional systems.}
\end{table*}

\section{Two-dimensional phase-space}

We now consider the effect of an additional component, not
explicitly coupled to the scalar field, but gravitationally coupled
via the Friedmann constraint for the Hubble expansion:
\be
\label{Friedmann2}
H^2 = {\kappa^2\over3} \left( {1\over2}\dot\phi^2 +
V + \rho_\gamma \right) \ .
\ee
where $\rho_\gamma$ obeys the continuity equation
\be
\label{continuity}
\dot\rho_\gamma = -3H(\rho_\gamma+P_\gamma) \,.
\ee
and the scalar field obeys the same Klein-Gordon equation~(\ref{KG}).

We will consider the simplest case of a barotropic fluid, with
pressure $P_\gamma=(\gamma-1)\rho_\gamma$ where $\gamma=$constant and
$0<\gamma<2$ for conventional fluids (such as dust, $\gamma=1$, or
radiation, $\gamma=4/3$).  
In this case we again have an autonomous system~\cite{CLW98} in terms
of dimensionless variable $x$ and $y$ defined in Eq.~(\ref{defxy}) and
a new dimensionless variable $w$
\be
w \equiv {\kappa\sqrt{\rho_\gamma}\over\sqrt{3}H} \,.
\ee

The evolution equations (\ref{KG}) and (\ref{continuity}) can then be
written as an autonomous system~\cite{WE,CLW98}:
\bea
\label{eomx2}
x' & = & -3x\left(1-x^2-{\gamma\over2}w^2\right)
 \pm \lambda \sqrt{{3\over2}} y^2
 \ , \\
\label{eomy2}
y' & = & y \left( 3x^2 + {3\gamma\over2}w^2 - \lambda \sqrt{{3\over2}}
  x \right) \ , \\
\label{eomw}
w' &=& {3\over2} w \left( -\gamma + 2x^2 + \gamma w^2 \right) \ ,
\eea
where as before a prime denotes a derivative with respect to $N$.

The Friedmann constraint (\ref{Friedmann2}) takes the form
\be
x^2 \pm y^2 + w^2 = 1 \,,
\ee
and reduces the system to a two-dimensional phase-space corresponding to
the unit sphere for $V\geq0$ or hyperboloid for $V\leq0$.
Note that $w=0$ remains an invariant one-dimensional subspace.
We will restrict our analysis to expanding cosmologies, $H>0$,
corresponding to the upper-quadrant $y\geq0$ and $w\geq0$, noting that
the system is symmetric under $t\to-t$, $H\to-H$, $y\to-y$ and $w\to-w$.

The same system of equations can also be
used to describe Bianchi type~I cosmologies \cite{vdH}, where the
shear appears in the constraint equation~(\ref{Friedmann2}) like a
fluid with $\gamma=2$, or spatially curved FRW models \cite{vdH2},
where the curvature appears in the Friedmann equation like a fluid
density with $\gamma=2/3$ (and $\rho_\gamma<0$, for positive curved
space, or $\rho_\gamma>0$, for negative curvature). In particular the
stability analysis for the barotropic fluids with $\gamma=2$ or
$\gamma=2/3$ is also applicable to scalar fields with exponential
potentials in Bianchi type~I or curved FRW models, where the flat FRW
models are an invariant one-dimensional subspace.

\subsection{Critical points}

Critical points correspond to fixed points $(x_i,y_i,w_i)$ where
$x'=0$, $y'=0$ and $w'=0$, and these are self-similar solutions
with, for example,
\be
\label{dotH2}
{\dot{H}\over H^2} = - 3x^2 - {3\gamma\over2}w^2 \,,
\ee
This corresponds to a power-law solution for the scale
factor
\begin{equation}
\label{powerlaw2}
a \propto |t|^p \,, \quad {\rm where}\ p = {2\over6x_i^2+3\gamma w_i^2} \,.
\end{equation}

The system~(\ref{eomx2}), (\ref{eomy2}) and (\ref{eomy}) has at most
five fixed points given by the vacuum/scalar field solutions $A_\pm$,
$B$, and two new points:
\begin{itemize}
\item[$C$]
A {\em fluid-dominated} solution
\be
x_C = 0 \,, \quad y_C=0 \,,\quad w_C=1 \,,
\ee
always exists, for any form of potential, corresponding to a power-law
solution given by Eq.~(\ref{powerlaw2}) and $p=2/3\gamma$.
\item[$D$]
A {\em fluid-potential-kinetic-scaling} solution exists for
steep positive potentials:
\begin{eqnarray}
x_D = \sqrt{3\over2}{\gamma\over\lambda}  \,,
 \qquad
y_D = \sqrt{\pm {3\over2}{(2-\gamma)\gamma\over\lambda^2}} \,,
\nonumber \\
w_D = 1-{3\gamma\over\lambda^2} \,.
\end{eqnarray}
This solution only exists for potentials that are both positive
($V>0$) and sufficiently steep ($\lambda^2>3\gamma^2$). The power-law
exponent in Eq.~(\ref{powerlaw2}), $p=2/3\gamma$, is identical to that
of the fluid-dominated solution and depends only on the barotropic
index and is independent of the slope of the potential.  This is the
classic example of `tracking' behaviour which occurs for a wide
variety of steep positive potentials~\cite{Wetterich,RP}.
\end{itemize}

\subsection{Stability}

The stability of $A_\pm$ and $B$ with respect to perturbations in the
$w=0$ sub-space is unchanged, yielding one eigenmode and eigenvalue
as discussed in sect.~\ref{1Dstability}. Introducing fluid perturbations
yields a second eigenmode $w'=3(2x_i^2-\gamma)w/2$:
\begin{itemize}
\item[$A_\pm$] Fluid perturbations evolve as $w\propto e^{mN}$ where
  the eigenvalue $m=3(2-\gamma)/2$. Thus the kinetic-dominated
  solutions are always unstable to fluid perturbations for $\gamma<2$.
  For a stiff fluid with $\gamma=2$ (or Bianchi type~I shear) the
  kinetic-dominated solutions are marginally stable.
\item[$B$] Fluid perturbations evolve as $w\propto e^{mN}$ where the
  eigenvalue $m=(\lambda^2-3\gamma)/2$. Thus the kinetic-potential
  scaling solution is stable with respect to fluid perturbations for
  sufficiently flat potentials, $\lambda^2<3\gamma$, but unstable for
  $\lambda^2>3\gamma$. In particular, the kinetic-potential scaling
  solutions in spatially-flat FRW models are stable with respect to
  spatial curvature for $\lambda^2<2$, yielding with power-law
  inflation for positive potentials~\cite{LM85}.
\end{itemize}
The two new points also have two eigenmodes:
\begin{itemize}
\item[$C$]
  The fluid-dominated solution has two eigenmodes $x'=-3(2-\gamma)x/2$
  and $y'=3\gamma y/2$. Thus the solution is stable to kinetic energy
  perturbations for $\gamma<2$, but unstable to potential energy
  perturbations for $\gamma>0$. Thus for $0<\gamma<2$ it is a saddle
  point and hence unstable to generic perturbations.
\item[$D$]
  Because point D can only exist for positive potential the
  phase-space reduces to that studied in Refs.~\cite{CLW98} where
  it was shown that this scaling solution is always the late-time
  stable attractor when it exists.
\end{itemize}

\subsection{Behaviour at infinity}

For positive potentials the one-dimensional system is compact, but for
$V\leq0$ the trajectories can go out to infinity.
We have
\be
\left( {x\over y} \right)^\prime = -3{x\over y}
 + \sqrt{3\over2}\lambda y \left( {x^2\over y^2} - 1 \right) \,,
\ee
and hence as $y\to\infty$ we find that generic solutions approach
$x/y\to-1$.

The remaining equations can then be written as
\be
(x^2+{\gamma\over2}w^2)' \to 3(x^2+{\gamma\over2})^2
\ee
so trajectories approach infinity with $x^2+(\gamma/2)w^2=
1/6(N_*-N)$. Equation~(\ref{dotH2}) can be integrated to obtain a
solution for the scale factor in terms of proper time
\be
N_*-N \propto (t-t_*)^2 \,,
\ee
showing that the behaviour at infinity always corresponds to
recollapse, i.e., a maximum of the expansion at finite time $t_*$.
Thus expanding solutions in the upper quadrant ($w,y>0$) are linked to
collapsing solutions in the lower quadrant ($w,y<0$) at infinity. When
$H$ changes sign, $x$, $y$ and $w$ all change sign.

\subsection{Qualitative evolution}

\begin{figure}[t]
\centering
\leavevmode\epsfysize=6cm \epsfbox{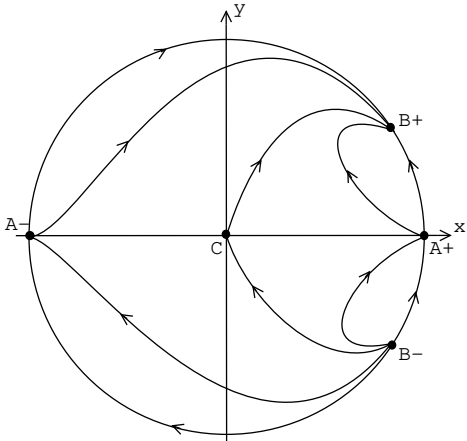}\\
\caption[hyperbsteep]{\label{discflat} Two-dimensional phase-space
  for flat positive potentials, $\lambda^2<3\gamma$.}
\end{figure}

\paragraph{Flat positive potentials ($V>0$, $\lambda^2<3\gamma$)}.
Four critical points exist in the upper quadrant $y\geq0$. Points
$A_+$ and $A_-$ are unstable repellors, $C$ is the unstable fluid
dominated solution, while $B$ is the stable late-time attractor.
Generic solutions begin at one or other of the kinetic dominated
solutions ($A_+$ or $A_-$) and approach the potential-kinetic solution
($B$) at late times. See Figure~\ref{discflat}. They can approach
arbitrarily close to the fluid dominated saddle point before reaching
the potential-kinetic solution.

\begin{figure}[t]
\centering
\leavevmode\epsfysize=6cm \epsfbox{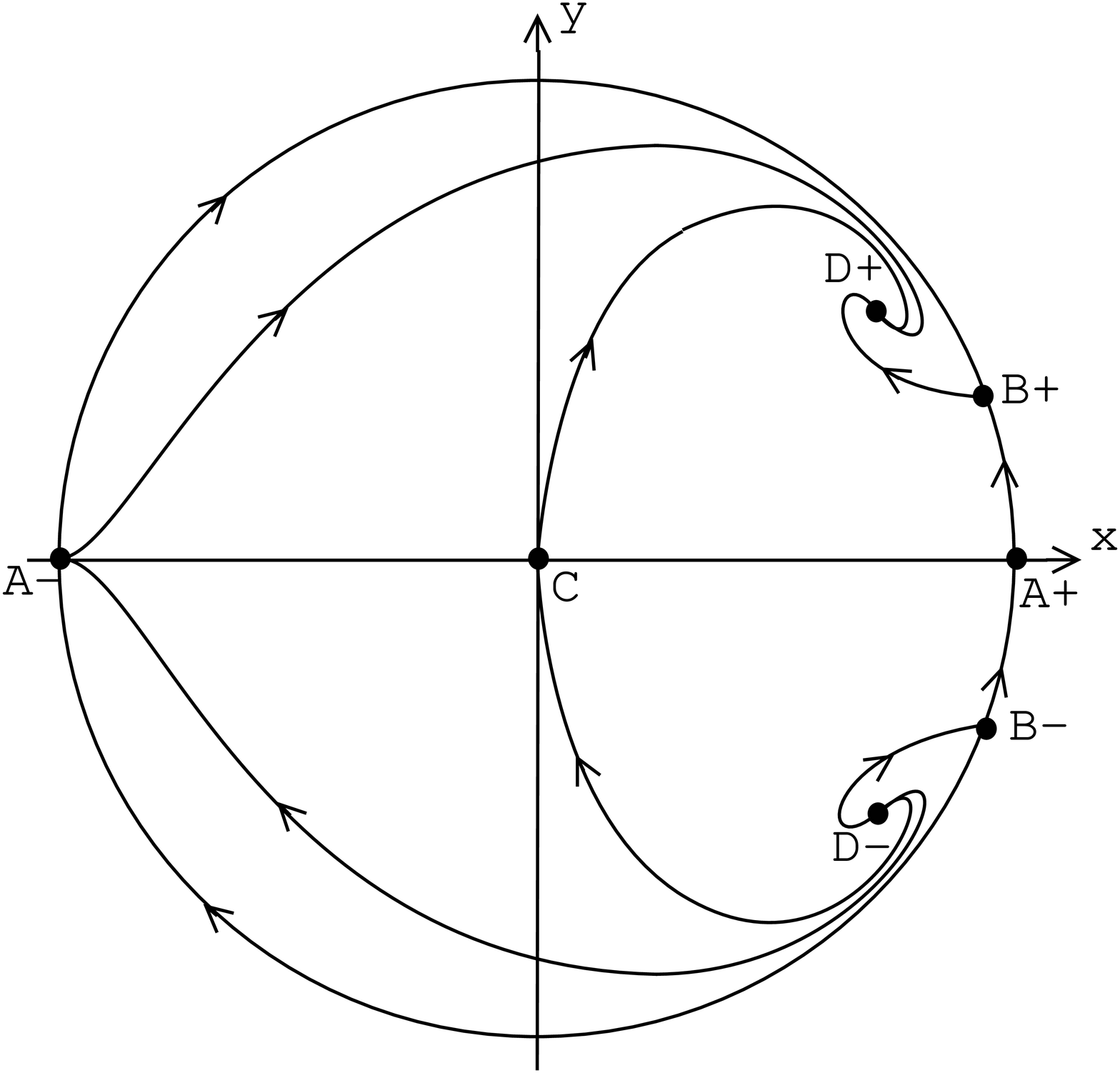}\\
\caption[hyperbsteep]{\label{discmed} Two-dimensional phase-space
  for intermediate positive potentials, $3\gamma<\lambda^2<6$.}
\end{figure}

\paragraph{Intermediate positive potentials ($V>0$,
  $3\gamma<\lambda^2<6$)}. All five critical points exist for
$y\geq0$. Generic solutions begin kinetic dominated ($A_+$ or $A_-$)
and may approach the fluid dominated ($C$) or potential-kinetic ($B$)
saddle points before going to the fluid-potential-kinetic scaling
solution ($D$) at late times. See Figure~\ref{discmed}.

\begin{figure}[t]
\centering
\leavevmode\epsfysize=6cm \epsfbox{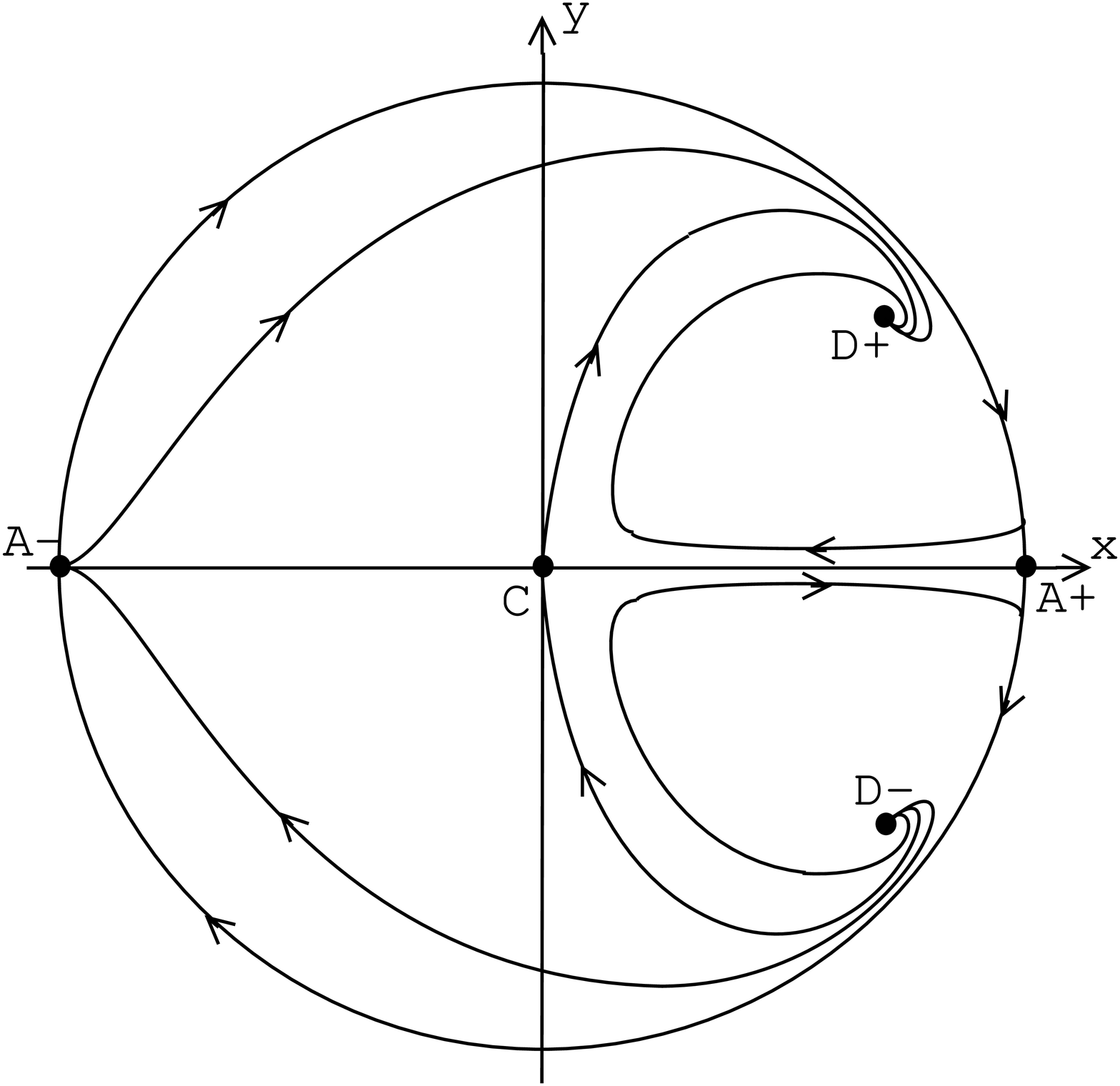}\\
\caption[hyperbsteep]{\label{discsteep} Two-dimensional phase-space
  for steep positive potentials, $\lambda^2>6$.}
\end{figure}

\paragraph{Steep positive potentials ($V>0$, $\lambda^2>6$)}. Four
critical points exist. Points $A_+$ and $A_-$ are unstable repellors,
$C$ is the unstable fluid dominated solution, while $D$ is the stable
late-time attractor. Solutions can begin in the kinetic dominated
solutions and may approach the fluid dominated saddle point ($C$)
before going to the fluid-potential-kinetic scaling solution ($D$) at
late times. See Figure~\ref{discsteep}. Those that start out kinetic
dominated can approach the scaling solution via the fluid dominated
solution.

\begin{figure}[t]
\centering
\leavevmode\epsfysize=6cm \epsfbox{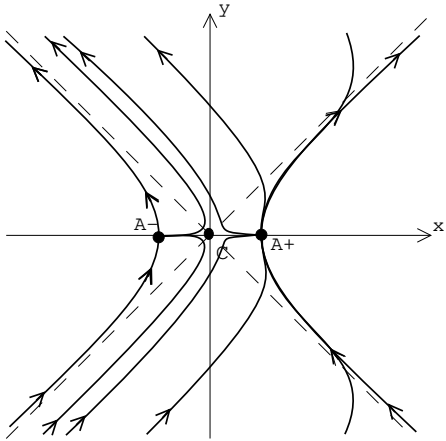}\\
\caption[hyperbsteep]{\label{hypflat} Two-dimensional phase-space
  for flat negative potentials, $\lambda^2<6$.}
\end{figure}

\paragraph{Flat negative potentials ($V<0$, $\lambda^2<6$)}. Three
critical points exist. Points $A_+$ and $A_-$ are unstable repellors
and $C$ is the unstable fluid dominated solution.  Generic solutions
begin in one of the kinetic dominated solutions ($A_+$ or $A_-$) and
go to infinity along the line $x=-y$, before recollapsing towards the
kinetic dominated collapse solution $A_+$. See Figure~\ref{hypflat}.
Expanding cosmologies that start out kinetic dominated can reach
arbitrarily close to the fluid-dominated saddle point before
recollapsing. Generic solutions have finite lifetime before
recollapsing to a singularity.

\begin{figure}[t]
\centering
\leavevmode\epsfysize=6cm \epsfbox{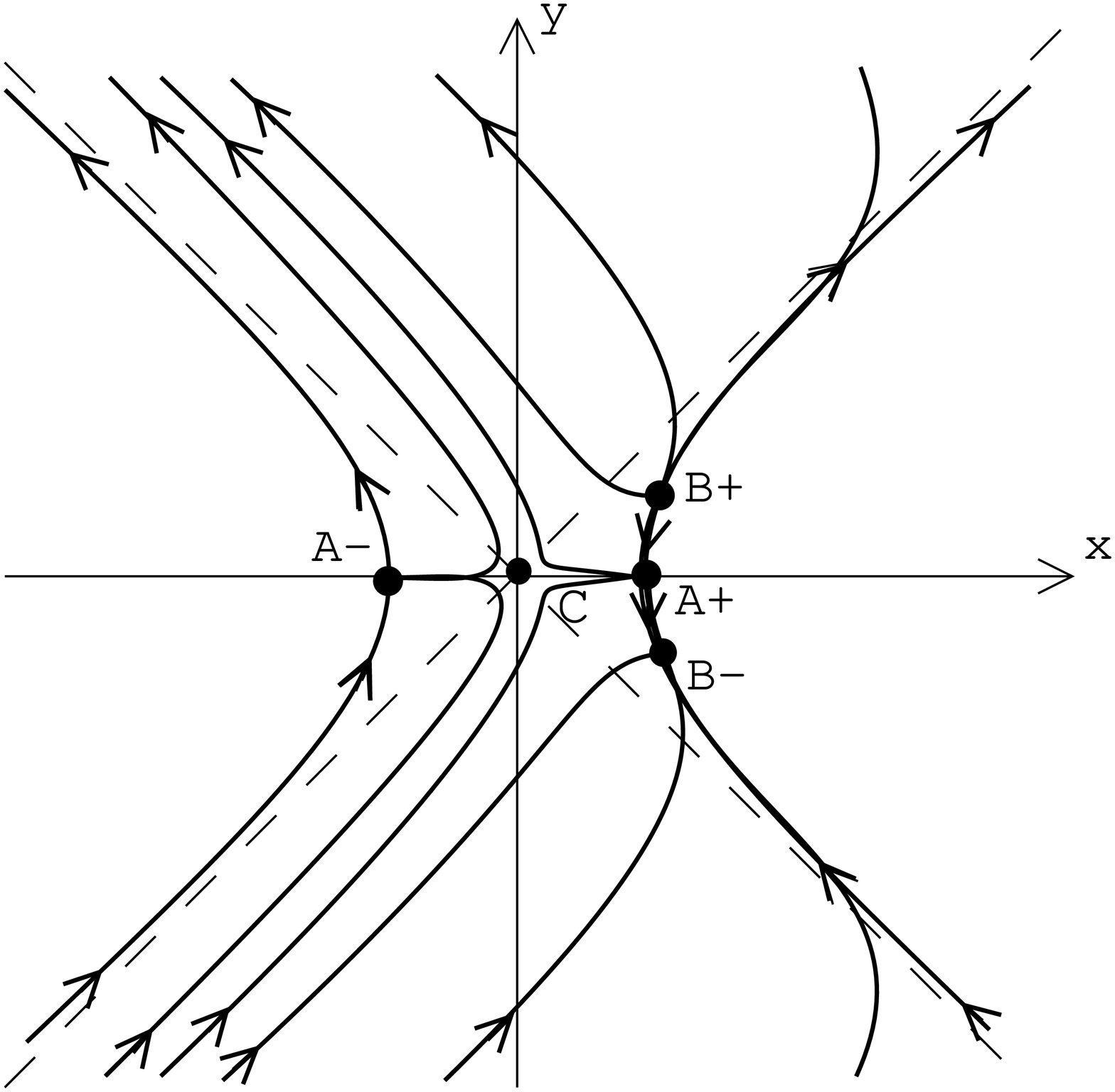}\\
\caption[hyperbsteep]{\label{hypsteep} Two-dimensional phase-space
  for steep negative potentials, $\lambda^2>6$.}
\end{figure}

\paragraph{Steep negative potentials ($V<0$, $\lambda^2>6$)}. Four
critical points exist and all are unstable. Generic solutions begin
either kinetic dominated ($A_-$ for $\dot\phi<0$) or kinetic-potential
scaling ($B$ for $\dot\phi>0$) and go to infinity along the line
$x=-y$, before recollapsing towards the kinetic-potential collapse
solution ($B$). See Figure~\ref{hypsteep}. If an expanding cosmology
begins in the kinetic dominated solution ($A_-$) it may approach the
fluid dominated saddle point ($C$) before going to infinity. If an
expanding cosmology begins in the potential-kinetic solution ($B$) it
may approach both the kinetic dominated saddle point ($A_+$) and the
fluid dominated saddle point ($C$) before going to infinity. Generic
solutions have finite lifetime before recollapsing to a singularity.

\section{Discussion}

We have presented a phase-space analysis of the qualitative evolution
of cosmological models with a scalar field, $\phi$, with positive or
negative exponential potential, $V\propto \exp(-\lambda\kappa\phi)$.
For positive potentials that are sufficiently flat ($\lambda^2<6$)
there is a self-similar scaling solution where potential and kinetic
energies remain proportional. This is the basis of the well-known
power-law inflationary solutions~\cite{LM85} with $a\propto t^p$ and
$p>1/3$.  There is no such scaling solution for flat negative
potentials, but sufficiently steep ($\lambda^2>6$) negative potentials
do have a self-similar power-law solution with $p<1/3$.

It has recently been found in the case of polynomial
potentials~\cite{Kofmanetal} that expanding cosmologies with a
negative potential always recollapse.  By contrast, we find a
expanding solutions with a sufficiently steep ($\lambda^2>6$) negative
potential that start close to the scaling solution at early times but
approach the kinetic-dominated solution at late times where the
potential is negligible and expansion can continue indefinitely.
Cosmologies with flat ($\lambda^2<6$) negative potentials do always
recollapse.

The scaling solution is always the late-time attractor for flat
positive potentials in an expanding universe, but the scaling solution
with steep negative potentials is always unstable in an expanding
universe. Conversely the kinetic-dominated solution with stiff
equation of state ($P_\phi=\rho_\phi$) is always the attractor
approaching a big crunch singularity in a collapsing universe with
positive potential, but for steep negative potentials the attractor
solution is the scaling solution with an ultra-stiff
($P_\phi>\rho_\phi$) equation of state. This is quite contrary to the
usual expectation that any scalar field potential becomes negligible
as the singularity is approached. For steep negative potentials the
isotropic ultra-stiff scaling solution remains the attractor even in
the presence of conventional matter perturbations, spatial curvature
and anisotropic shear.

It is interesting to apply these results to recently proposed
cosmological models based on exponential potentials in a `pre big
bang' phase.
The ekpyrotic model~\cite{ekpyrotic,pyro} has a steep negative
potential ($\lambda^2\gg6$). Thus the kinetic-potential scaling
solution is the attractor solution approaching the big crunch
singularity in a collapsing universe.  It is stable to both
conventional matter perturbations, curvature and shear perturbations.
This echoes recent results found for anisotropic cosmologies in
Randall-Sundrum-type brane-world scenarios~\cite{ColeyBW}, but is in
contrast to the pre big bang scenario for a massless dilaton
field~\cite{pbb} which is only marginally stable to anisotropic
shear~\cite{hair}.

Another topical example is Finelli and Brandenberger's collapsing
model~\cite{FinBra} that gives a scale-invariant spectrum of comoving
curvature perturbations during the collapse phase. This corresponds to
a kinetic-potential scaling solution with $p=2/3$, i.e.,
$\lambda^2=3$, so requires a flat, positive potential. Thus this
solution is unstable in a collapsing universe where generic solutions
approach the kinetic-dominated regime.

\acknowledgments

IPCH is supported by an EPSRC studentship. DW is supported by the
Royal Society.


\end{document}